\documentclass[onecolumn,amsmath,superscriptaddress,pra,aps,10pt]{revtex4-1}
\usepackage{graphicx}
\usepackage{bm,verbatim}
\usepackage{color}
\usepackage{float}
\usepackage{amsmath}
\usepackage{lineno}

\begin{document}

\title{Metasurface electron optics in graphene}

\author{Ruihuang Zhao}
\thanks{These authors contributed equally to this work.}
\affiliation{State Key Laboratory of Precision Spectroscopy, School of Physics and Electronic Science, East China Normal University, Shanghai 200062, China\\}
\author{Pengcheng Wan}
\thanks{These authors contributed equally to this work.}
\affiliation{State Key Laboratory of Precision Spectroscopy, School of Physics and Electronic Science, East China Normal University, Shanghai 200062, China\\}
\author{Ling Zhou}
\affiliation{State Key Laboratory of Precision Spectroscopy, School of Physics and Electronic Science, East China Normal University, Shanghai 200062, China\\}
\author{Di Huang}
\affiliation{State Key Laboratory of Precision Spectroscopy, School of Physics and Electronic Science, East China Normal University, Shanghai 200062, China\\}
\author{Haiqin Guo}
\affiliation{State Key Laboratory of Precision Spectroscopy, School of Physics and Electronic Science, East China Normal University, Shanghai 200062, China\\}
\author{Hao Xia}
\affiliation{State Key Laboratory of Precision Spectroscopy, School of Physics and Electronic Science, East China Normal University, Shanghai 200062, China\\}
\author{Junjie Du}
\email{phyjunjie@gmail.com}
\affiliation{State Key Laboratory of Precision Spectroscopy, School of Physics and Electronic Science, East China Normal University, Shanghai 200062, China\\}

\date{\today }

\begin{abstract}
For electron optics in graphene, the propagation effect has so far been the only physical mechanism available.
The resulting electron-optics-based components are large in size and %
operate at low temperatures to avoid violating the ballistic transport limits.
In this paper, Dirac fermion metasurfaces, electronic counterparts of optical metasurfaces,  
are introduced for graphene electronics.
By a metasurface, formally a linear array of gate-bias-controlled circular quantum dots,
the wavefront of electron beams can be shaped within a one-quantum-dot-diameter distance, far below the ballistic limits at room temperature.
This provides opportunities to create electron-optics-based devices that operate under ambient conditions.
Moreover, unlike optical metasurfaces, Dirac fermion metasurfaces have near-perfect operating efficiencies and their high tunability allows for free and fast switching among functionalities.
The concept of metasurface electron optics
might open up a promising avenue for improving the performance of quantum devices in Dirac fermion materials.
\end{abstract}

\maketitle

\affiliation{State Key Laboratory of Precision Spectroscopy, School of Physics and Electronic Science, East China Normal University, Shanghai 200062, China\\}




Low energy graphene electrons behave like light because of their light-like dispersion and the ballistic transport\cite{graphene}.
Electron optics in semiconductor structures can be naturally extended to graphene. 
Both naturally-occurring and non-naturally-occurring optical phenomena
such as the Goos-Hanchen shift\cite{goosh}, self-collimation\cite{collimation2}, whispering-gallery modes\cite{WGM1,WGM2,WGM3,WGM4,WGM5}, and negative-index\cite{negative1,negative2,negative3,negative4} and zero-index\cite{zero} behaviors,
have been reproduced by graphene electrons.
Accordingly, various optics-inspired functional units,
such as two-dimensional electron microscopes\cite{microscope1,microscope2}, quantum switches\cite{switch1,switch2,switch3}, Fabry-P\'{e}rot cavities\cite{Fabry}, electron waveguides\cite{waveguid1,waveguid2,waveguid3}, splitters\cite{splitter1,splitter2} and Veselago lenses\cite{negative2,negative3,negative4}, have been demonstrated.
However, these remarkable achievements were made by following procedures used for bulk optical materials 
where wavefront shaping is accomplished via light propagation over a distance much larger than the wavelength.
The propagation effect dictates that the optics-inspired electronic components are large in size.
But, even if the large size of devices can be tolerated,
the required long propagation distances often challenge the ballistic transport limits.
To avoid violating these ballistic transport limits,
these components have been designed to operate at low temperature 
since the mean free path of graphene electrons, $l=\mu \sqrt{\rho/\pi} h/2e$, which determines the ballistic transport limit, becomes larger at low temperature with enhanced carrier mobility $\mu$\cite{temperature}.

The emergence of optical metasurfaces\cite{metasurfacesYu,metasurfacesNi} and metagratings\cite{DuPRL11,DuPRL13} opens the door to flat optics technology characterized by a single layer of nanoparticles.
Wavefront shaping in optical metasurfaces is achieved
over the scale of the free-space wavelength or on a smaller scale
by introducing abrupt changes in phase or amplitude, following the generalized Snell's law\cite{metasurfacesYu,flatoptics}.
This is fundamentally different from conventional diffractive optics
based on the propagation effect.
A single-particle-layer material not only greatly simplifies the fabrication process 
and lowers the loss in contrast to bulk materials,
but also can mould optical wavefronts into shapes that are designed at will\cite{flatoptics}.
Inspired by the compactness features and the remarkable capabilities of wavefront engineering,
we explore the possibility of realizing Dirac fermion metasurfaces in graphene,
aiming to develop ultrasmall optics-inspired transistors that can operate under ambient conditions. 
In addition to the inherent advantages of metasurfaces such as compactness, low loss and easy-fabrication,
it is shown that Dirac fermion metasurfaces also possess other remarkable properties such as near-perfect operating efficiency and
high tunability, properties that are difficult to achieve in their optical counterparts.
Also worth noting is that Dirac fermion metagratings have been explored in graphene and successfully produces unit-efficiency beam deflection
through a near-$180^0$ angle over a distance much smaller
than the electron wavelength\cite{metagrating}.

\section*{Results and Discussion}
Formally, a Dirac fermion metasurface is 
a linear array of gate-bias-controlled circular quantum dots (QDs).
The QDs
are responsible for providing the required phase response in constructing a constant gradient of phase jump.
The phase response capability of QDs
is related to their ``refractive index''
which is defined as
$n_s=(E-V_s)/E$
where $E$ is the incident energy and $V_s$ the applied bias\cite{Heinisch}.
For the electron scattering problem
in the single valley case (see Supplemental Material),
the low-energy electron dynamics can be described by the Dirac-Hamiltonian\cite{Heinisch,Katsnelson,Cserti,Pieper1,Novikov,Ostrovsky,Hentschel}
\begin{equation} \label{hamilton}
H=-i\hbar v_F\nabla\sigma+V_s\Theta(R_s-r),
\end{equation}
which is analogous to the light scattering problem of an infinite dielectric cylinder.
The Mie scattering method used widely in optics 
is applicable to graphene electrons; 
some of the results predicted by the method
have been experimentally verified\cite{E-Mie}.
QDs with radius $R_s$ are denoted as a step potential in Eq.~(\ref{hamilton}) with the Heaviside step function $\Theta(R_s-r)$.
The potential is smooth on the scale of the graphene's
intrinsic lattice constant but sharp on the scale of de Broglie wavelength,
so the intervalley scattering is negligible.
Meanwhile, theoretical studies have also shown that the QDs
have nearly the same electron scattering behaviors\cite{Pieper2} 
for the gradual transition of potentials smaller than $0.5R_s$. 

As is well known, the behavior of a wave in a metasurface follows the generalized Snell's law of refraction\cite{metasurfacesYu}
\begin{equation}
n_t\sin\theta_t-n_i\sin\theta_i=\frac{1}{k_0}\frac{d\Phi}{dx},
\end{equation}
where $k_0$ is the magnitude of the free space wavevector, $\theta_i$ and $\theta_t$ are, respectively, the angle of incidence and refraction,
and $n_i$ and $n_t$ are the respective ``refractive indices'' of media on the incident and transmission sides of the metasurface.
The phase gradient $d\Phi/dx$ implies an effective wavevector (equivalently, an effective momentum) along the interface
that is produced and is imparted to the transmitted and reflected electrons.
Thus the transmitted and reflected electron beams can be deflected through arbitrary angles,
depending on the direction and magnitude of the phase gradient.
The phase gradient is created within a unit cell consisting of several QDs
to which the linearly increasing biases $V_s$ are respectively applied.
The variation of bias indicates the difference in ``refractive index" between the QDs 
and thus the difference in phase response of electron waves.
Figure 1 illustrates a linear phase distribution of the scattering fields in a unit cell composed of ten QDs.
Throughout this paper, the energy of the incident electron beams is chosen to be $E=65.82$ meV
and the radius of the QDs to be 5 nm, with a spacing $d=14.6$ nm between them.
So the period of the metasurfaceis in Fig.~1 is $\Gamma=9d=131.4$ nm;
in this figure the biases are given above each scattering field plot.
Figure 1 shows that a complete phase coverage from 0 to $2\pi$ is obtained with an approximately constant phase
difference $\Delta\phi=\pi/5$ between neighbors.
Thus the magnitude of the introduced wavevector in the x direction is $k_x^{add}=d\Phi/dx=2\pi/\Gamma=0.048$/nm.
When a normal-incidence electron beam impinges on the metasurface,
the transmitted beam will be bent at an angle $\theta_{calc}=\arctan (k_x^{add}/k_0)$,
where $k_0$ is the magnitude of the free space wavevector with $n_i=n_t=1$.
This is simulated in Fig.~2(a), and the travel direction of the transmitted beam agrees well with the calculated bending angle $\theta_{calc}=25^\circ$.
Note that the scattering field of each QD in Fig.~1 is calculated by considering
the inter-QD coupling interaction 
and employing multiple scattering theory\cite{metagrating,effective,coup}(see also Supplemental Material).
This ensures that the constant gradient of phase jumps really exists in the metasurface
since the scattering field of an isolated QD may be very different from that of the same QD in a linear array.



A unit cell which covers the entire 0-2$\pi$ range can also be composed of different numbers of QDs
by adjusting only the bias while keeping the array invariant.
We denote the number of the QDs in a unit cell by $m_{unit}$.
A decrease of $m_{unit}$ implies a larger phase gradient and also a larger introduced wavevector.
Figures 2(b)-(f) show the bending of the electron beams by metasurfaces with $m_{unit}$=9, 8, 7, 6, and 5.
The bending angle increases gradually from Fig.~2(b) to (f).
In addition, the Dirac fermion matasurfaces show a remarkable property in Fig.~2,
that electron waves can be bent with nearly perfect efficiency.
This is strikingly different from optical metasurfaces,
in which an efficiency near 100$\%$ is difficult to achieve.
It shows that electrons can more easily react to the lateral momentum introduced by the phase gradient than can photons.
Moreover, comparing the panels in Fig.~2 shows that
the efficiency is closer to 100$\%$ in the case of the longer unit cells
because electrons have more opportunities to react to the introduced lateral momentum.
Finally, perfect efficiency cannot be divorced from the successful suppression of reflection in the metasurfaces (see
Supplemental Material Figure S2).

To verify that the beam bending at various $\theta_{actu}$ in Fig.~2 can be well explained by the introduced wavevectors due to phase gradient,
we made a comparison between the two for all the cases in Table I
where $\tan\theta_{calc}=k_x^{add}/k_0$ and $\tan\theta_{actu}$ are given.
Here $k_x^{add}=2\pi/\Gamma=2\pi/m_{unit}d$ and $\theta_{actu}$ is directly read from Fig.2.
The calculated bending angles agree well with the actual ones for $m_{unit}$ between 7 and 10,
but there are distinct deviations for $m_{unit}$=6 and 5.
To make matters worse, a weak beam is transmitted to the left of the normal in the later two cases.
In an effort to find the causes of these deviations,
we examine the phase response of each quantum dot for the case of $m_{unit}$=6, as shown Fig.3(a).
One can see that only the former five QDs contribute to the formation of the linear phase gradient,
whereas the QD with $V_s$=750 meV has a phase that goes against the linear gradient change.
Moreover, this QD and the next two QDs of $V_s$=680 meV and $V_s$=400 meV
together form a phase gradient increasing in the opposite direction; hence a left-oriented wavevector is produced, as shown in Fig.~3(a).
Thus a small portion of electrons will propagate on the left side of the normal in Fig.~2(e).
Accordingly, the magnitude of the introduced right-oriented wavenumber should be calculated in terms of the period $\Gamma'=4d=58.4$nm.
The new $\tan\theta_{calc}$ calculated by $k_x^{add}=2\pi/\Gamma'$ is equal to 1.075 and agrees well with $\tan\theta_{actu}$ in Table I.
Similarly, the deviation in the case of $m_{unit}$=5 in Table I has the same cause (see Supplemental Material Figure S1).

Exploring the causes of deviations can help to improve the efficiency of metasurfaces.
We note that the electron scattering of the QD of $V_s$=750 meV is weak in contrast to other QDs in the unit cell.
So the scattering of the other QDs will hardly be impacted if this QD is removed from the unit cell.
Such a unit cell is schematically shown in the lower panel in Fig.~3(b),
and the phase distribution is given in Fig.~3(c).
We see that the opposite phase gradient is eliminated and thus only a right-oriented wavevector is introduced.
The electron density distribution displayed in Fig.~3(d) shows that
the electron beam bends to the right side of the normal with near-unit efficiency
when the QD of $V_s$=750 meV is removed.

Since two equal and oppositely directed phase gradients
represent left- and right-oriented equal-magnitude wavevectors,
we can introduce them simultaneously in a metasurface to design an ultrathin electron splitter.
One simple route for splitters is to achieve a right-oriented wavevector by the unit cells in the right half of the array
and a left-oriented wavevector by the unit cells in the left half.
This can be implemented by applying the biases enhanced from left to right to the QDs in the unit cells in the right half,
as shown in Fig.~2,
and the same biases but enhanced in the opposite direction in the left half.
The impinging beam is split into two sub-beams
at various angles to each other, 
as demonstrated in Figs.~4(a)-(f). 
We see that the beams are split with nearly perfect efficiency again
and the splitting ratio is 50-50 in all cases.

Three points are worth emphasizing in the model.
First, all the results reported in this paper
are obtained in the same linear array of QDs.
Specifically, the radius of the QDs and the spacing between them remain invariant in all simulations
and we only modulate the biases on the QDs to realize both beam bending and beam splitting at various angles.
The fast switching time of bias systems allows for high modulation efficiency.
Second,
near-perfect efficiency is obtainable, fundamentally different from optical counterparts.
The performance of optical metasurfaces is subject to the intrinsic nature of light.
The introduced momentum
through the phase gradient
cannot be perceived by all the photons due to the absence of interaction between photons.
Because electrons are distinct from photons, the Dirac fermion metasurfaces have near-perfect operating efficiency.
Third, the 5-nm radius QDs used in our simulations fall within current experimental manufacturing tolerance\cite{QDexperiment1,QDexperiment2,QDexperiment3}.
Very recently, even smaller circular QDs with atomically
sharp boundaries have been obtained in experiments\cite{QDexperiment1,QDexperiment2,QDexperiment3}.
The fabrication techniques of high-precision QDs
makes experimental realization of the metasurfaces feasible.

\section{CONCLUSIONS}

In summary, we have demonstrated theoretically the feasibility of realizing metasurfaces for graphene ballistic electrons.
A simple metasurface is a linear array of quantum dots (QDs) of the same radius.
Phase discontinuities, the essential ingredient of gradient metasurfaces, are acquired by applying difference biases to the QDs.
Following the generalized Snell's law, the metasurface imposes a control over electrons in a rather compact way
with wavefront shaping accomplished below the ballistic transport limit at room temperature.
Such metasurfaces dramatically reduce the size of electron-optics-based components
and enable them to get rid of the dependence on low temperature conditions.
The two kinds of transistors we demonstrated, beam benders and beam splitters, are achieved
in the same linear array of QDs and can be conveniently switched back and forth by tuning the biases applied to the QDs.
Dirac fermion metasurfaces represent a promising way to develop more practical and accessible electron optics technologies.

\newpage

\begin{table}[h]
\caption{\textbf{Comparison between the calculated and actual bending angles for various $m_{unit}$.}}
\setlength{\tabcolsep}{2.2mm}
\begin{tabular}{lcccccc}
  \hline
    $m_{unit}$ & 10 & 9 & 8 & 7 & 6 & 5 \\
  \hline
    $\tan\theta_{calc}$ & 0.480 & 0.535 & 0.600 & 0.715 & 0.861 & 1.074 \\
    $\tan\theta_{actu}$ & 0.468 & 0.526 & 0.620 & 0.760 & 1.000 & 1.700 \\
  \hline
\end{tabular}
\end{table}

\begin{figure*}[htbp]
\includegraphics[width=0.96\textwidth]{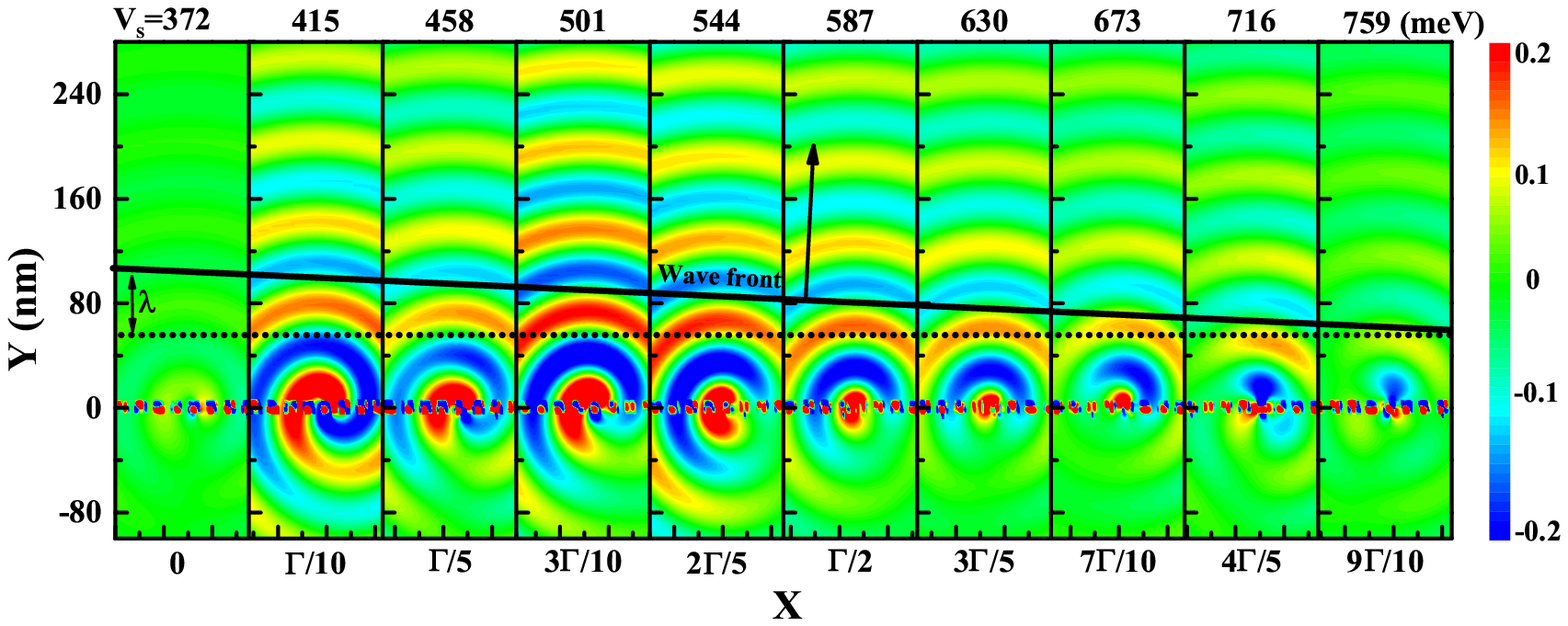}
\caption{\textbf{Formation of linear phase gradient.}
The scattering filed of the individual QDs constituting the unit cell of a metasurface.
The tilted black straight line is the envelope of the projections of the cylindrical waves scattered by the QDs.
A complete phase coverage from 0 to 2$\pi$ is shown with an approximately constant phase
difference $\Delta\phi=\pi/5$ between neighbors.
The number of QDs in the unit cell is $m_{unit}$=10 and the biases $V_s$ applied on each QD are given above each plot.
The inter-QD coupling interaction has been considered in this calculation.
}
\label{fig1}
\end{figure*}

\begin{figure*}[thbp]
\includegraphics[width=0.96\textwidth]{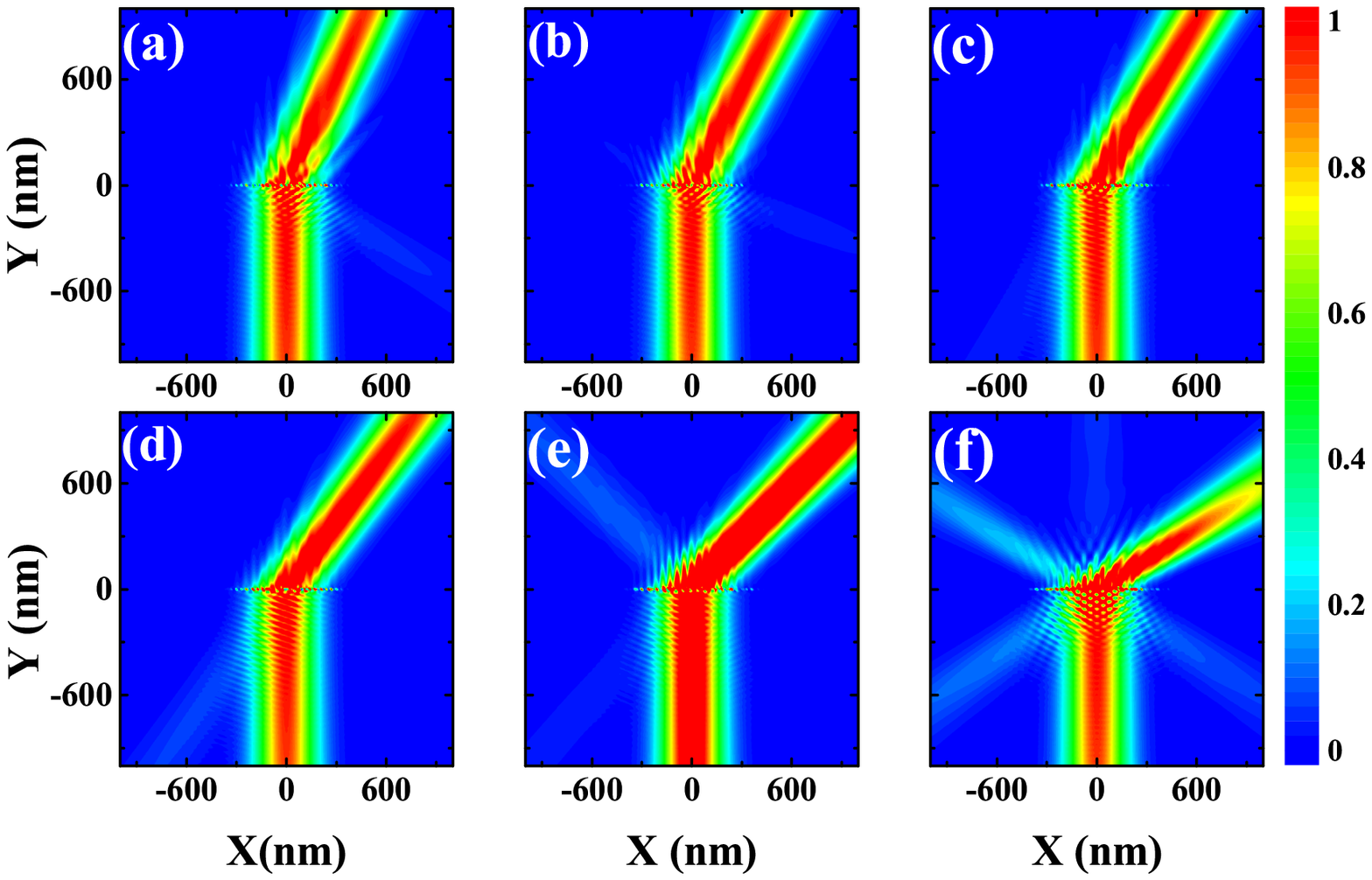}
\caption{\textbf{Simulation of beam bending.}
The bending of electron beams after passing through the metasurface composed of unit cells with the number of QDs $m_{unit}$
equal to (a) 10, (b) 9, (c) 8, (d) 7, (e) 6, and (f) 5.
Linearly increasing biases with constant gradient are applied to the QDs in the unit cells.
They are, respectively,
$V_s$=375, 420, 465, 510, 555, 600, 645, 690, and 735 meV for $m_{unit}$=9;
$V_s$=385, 440, 495, 550, 605, 660, 715, and 770 meV for $m_{unit}$=8;
$V_s$=385, 440, 495, 550, 605, 660, and 715 meV for $m_{unit}$=7;
$V_s$=400, 470, 540, 610, 680, and 750 meV for $m_{unit}$=6;
and $V_s$=390, 450, 510, 570, and 630 meV for $m_{unit}$=5.
}
\label{fig2}
\end{figure*}

\begin{figure*}[thbp]
\includegraphics[width=0.96\textwidth]{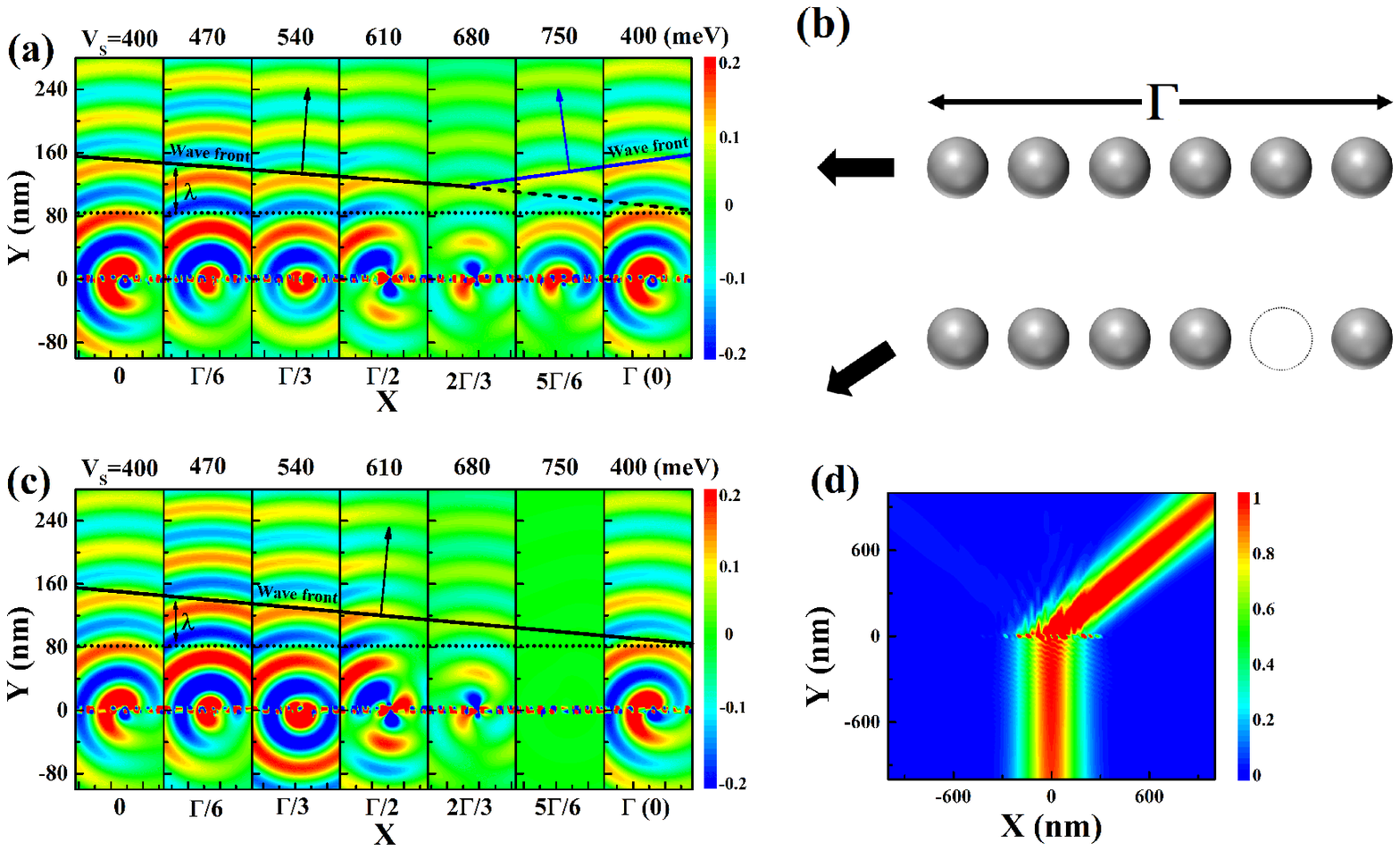}
\caption{\textbf{Illustration of the deviation at $m_{unit}=6$ in Table I.}
(a) The scattered filed of the individual QDs in the presence of all six QDs.
The tilted black and blue solid lines indicate, respectively, the desired and undesired phase gradients formed in the unit cell.
(b) Schematics of the unit cells in the presence of all the QDs in the upper panel and in the absence of the QD of $V_s=750$ meV in the lower panel.
(c) The scattered filed of the individual QDs with the QD of V=750 meV removed.
The phase gradient increasing in the opposite direction that appears in (a) is eliminated.
(d) The beam bending occurring with near-unit efficiency
when the QD of $V_s$=750 meV is removed, in sharp contrast to Fig.~2(e).
}
\label{fig3}
\end{figure*}

\begin{figure*}[thbp]
\includegraphics[width=0.96\textwidth]{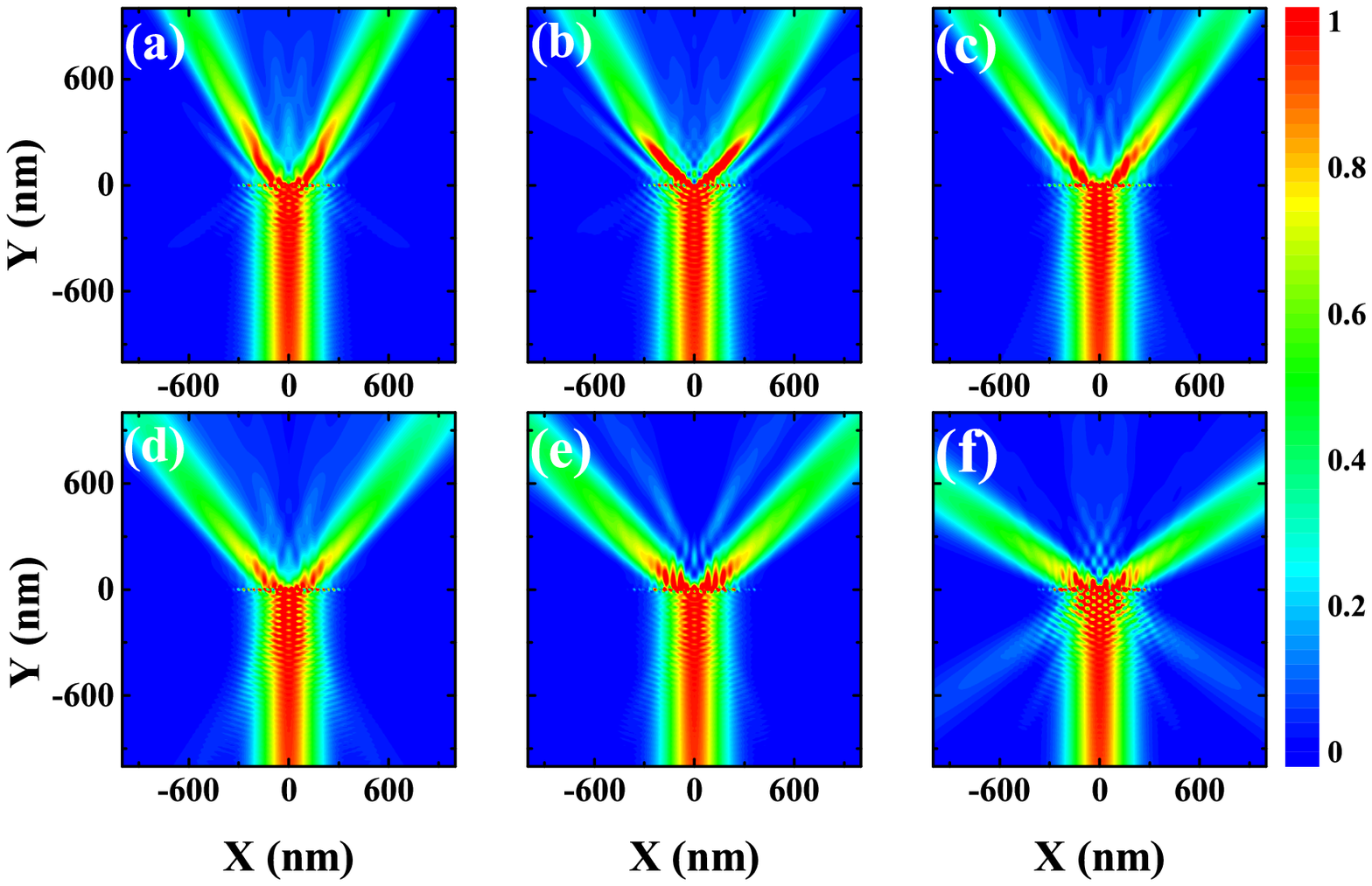}
\caption{\textbf{Simulation of beam splitting.}
The splitting of electron beams after passing through metasurface
composed of unit cells with the number of QDs $m_{unit}$ equal to
(a) 10, (b) 9, (c) 8, (d) 7, (e) 6, and (f) 5.
The unit cells in the right half of the metasurfaces in (a)-(f) are the same as those in Fig.2(a)-(f), respectively,
whereas the unit cells in the left half have the biases increasing in the opposite direction.
}
\label{fig4}
\end{figure*}

\end{document}